\documentclass{article}
\usepackage{spconf,amsmath,graphicx}

\usepackage{mwe} % to get dummy images
\title{Structural brain asymmetries in youths with combined and inattentive presentations of Attention Deficit Hyperactivity Disorder}
\name{Cintya Nirvana Dutta\textsuperscript{1,2}, Pamela K. Douglas\textsuperscript{2,3}, Hernando Ombao\textsuperscript{1}}

\address{\textsuperscript{1}Biostatistics Group, CEMSE, King Abdullah University of Science and Technology, Saudi Arabia
\\ \textsuperscript{2}SMST; Department of Computer Science; University of Central Florida, USA
\\ \textsuperscript{3}Psychiatry \& Biobehavioral Medicine, University of California, Los Angeles, USA}

\begin{document}
%\ninept
%
\maketitle
\begin{abstract}
Alterations in structural brain laterality are reported in attention-deficit/hyperactivity disorder (ADHD). However, few studies examined differences within presentations of ADHD. We investigate asymmetry index (AI) across 13 subcortical and 33 cortical regions from anatomical metrics of volume, surface area, and thickness. Structural T1-weighted MRI data were obtained from youths with inattentive ($n=64$) and combined ($n=51$) presentations, and aged-matched controls ($n=298$). We used a linear mixed effect model that accounts for data site heterogeneity, while studying associations between AI and covariates of presentation and age. Our paper contributes to the functional results seen among ADHD presentations evidencing disrupted connectivity in motor networks from ADHD-C and cingulo-frontal networks from ADHD-I, as well as new findings in the temporal cortex and default mode subnetworks. Age patterns of structural asymmetries vary with presentation type. Linear mixed effects model is a practical tool for characterizing associations between brain asymmetries, diagnosis, and neurodevelopment.

\end{abstract}
\vspace{-5pt}
\section{Introduction}
\label{sec:intro}
The Diagnostic and Statistical Manual of Mental Disorders (DSM-V) identifies three presentation types of attention-deficit/hyperactivity disorder (ADHD): predominant inattentive (ADHD-I), predominant hyperactive/impulsive (ADHD-H), and combined (ADHD-C) (DSM-V, 2013) \cite{american2013diagnostic}. The heterogeneity of ADHD features presents a challenge in the categorical grouping among individuals \cite{bush2010attention}. To complement the clinical diagnostic approach, a neurobiological framework of ADHD will provide opportunities for improved personalized treatment regimens for each presentation pathophysiology. Currently, there is little knowledge as to the neural mechanisms that drive each ADHD phenotypic profile, particularly in structural MRI results. Our goal is to fill this gap by addressing the differences in brain asymmetry among ADHD phenotypes using the anatomical metrics of volume, surface area, and thickness.

The latest findings do not ascertain the lack of differences in structural brain volume, surface area, or thickness among the ADHD presentations (for a review, see \cite{saad2020review}). Many voxel-based morphometry studies found no significant differences in gray matter volumes between ADHD-I and ADHD-C, \cite{vilgis2016global,carmona2009ventro,saad2017regional} and controls \cite{pineda2002statistical}. There were few studies with significant results, but different methodologies \cite{saad2020review}. Relative to ADHD-I and controls, ADHD-C had smaller voxel-wise hippocampal volumes \cite{al2018hippocampal}. Mixed findings are found when comparing one subtype to controls. ADHD-C children had greater volume in the hippocampus and left orbitofrontal region \cite{plessen2006hippocampus}. Reversal of asymmetry is found between ADHD-C (small left laterality) and healthy children (small right laterality) in putamen \cite{wellington2006magnetic}, suggesting brain asymmetry is an avenue to investigate.

Brain asymmetry is central to the brain's functional and structural organization \cite{toga2003mapping}. Measures of laterality, such as the asymmetry index (AI), provide a benchmark for evaluating the degree of subject-specific differences in laterality \cite{kurth201512,guadalupe2017human,kong2020mapping}. Differences in AI among structural MRI volume, surface area, and thickness measures have been found between typically developing (TD) and ADHD youths \cite{postema2020analysis,douglas2018hemispheric,dutta2020inter}. 
To the best of our knowledge, this is the first paper that addresses AI differences between ADHD presentations.

To address prior limitations in the literature, we introduce the following framework: (1.) We evaluate AI among volume, surface area, and thickness regions between two ADHD presentations (ADHD-C; ADHD-I) and typically developing youths; (2.) Using a linear mixed effect model, we used site as a random intercept to account for site heterogeneity (i.e., case recruitment, geographical location), similar to the dataset approach by \cite{postema2020analysis,kong2020mapping}; (3.) We examine the covariates of presentation, age, and their interaction. At present, there is no existing study that evaluates age as a factor that influences neural structures in ADHD presentations \cite{saad2020review}. This is important because the onset, severity, and persistence of ADHD presentations are suggested to differ with age \cite{faraone_attention-deficithyperactivity_2015}, and hence structural neurodevelopment in youths is addressed in this paper.
\begin{table}
  \centering
  \caption{Demographics for typically developing (TD), ADHD-I, and ADHD-C. We indicated total sample size, gender (Male-M / Female-F), handedness (Left-L / Right-R); and the mean and standard deviation for age and full IQ. The data sites were Peking University, Kennedy Krieger Institute (KKI), New York University (NYU) Child Study Center, and Oregon Health and Science University (OHSU).}\label{table:1}
  \bigbreak
 \scalebox{0.68}{
  \begin{tabular}{l|l|l|l|l|l}
    \textbf{Diagnosis}&\textbf{Measure}&\textbf{Peking}&\textbf{KKI}&\textbf{NYU}&\textbf{OHSU} \\
    \hline
    \textbf{TD}&Total&115&52&94&37 \\
    &M/F&70/45&29/33&45/49&15/22 \\
    &L/R&0/115&3/29&27/67&0/37 \\
    &Age&11.7 ± 1.7&10.3 ± 1.3&12/3 ± 3.2&8/9 ± 1.2 \\
    &Full IQ&118.1 ± 13.4&111.4 ± 10.6&110.7 ± 14.4&118.4 ± 12.3 \\
     \hline
    \textbf{ADHD-I}&Total&36&5&13&10 \\
    &M/F&30/6&4/1&7/6&5/5 \\
    &L/R&0/36&1/4&3/10&0/10 \\
    &Age&12.9 ± 2.0&11.3 ± 1.1&10.5 ± 2.6&8.9 ± 0.8 \\
    &Full IQ&101.2 ± 12.3&103.2 ± 12.7&107.8 ± 14.1&102.8 ± 14.8 \\
     \hline
    \textbf{ADHD-C}&Total&16&10&17&8 \\
    &M/F&16/0&4/6&13/4&8/0 \\
    &L/R&1/15&0/10&5/12&0/8 \\
    &Age&12.2 ± 1.5&10.4 ± 1.9&9.8 ± 2.3&8.5 ± 0.6 \\
    &Full IQ&110.4 ± 10.6&109.7 ± 16.4&105.1 ± 12.3&115 ± 9.7 \\
  \end{tabular}}
\end{table}

\section{Compliance with Ethical Standards}
\label{sec:data}
We obtained structural T-1 weighted MRI scans from the preprocessed ADHD-200 Repository through Athena pipeline \cite{bellec2017neuro}. This study was conducted on retrospective human subject data made available in open access by \cite{bellec2017neuro} and thus no ethical approval was required as stated by their license. There were a total of 1198 structural scans available. Our exclusion criteria included subjects with more than 1 scanning session ($n=430$), medicated/unknown medication status ($n=263$), prior history of other psychological or neurological disorder ($n= 82$), and any questionable quality control reported by the Athena pipeline ($n=10$). The demographics for subjects used in the analysis ($n=413$) are included in Table \ref{table:1}.

We gathered all preprocessed structural T-1 weighted brain scans from the Athena Pipeline in NITRC (NeuroImaging Tools and Resources Collaboratory) \cite{bellec2017neuro}. All information on the preprocessing steps can be viewed in \cite{bellec2017neuro}. Using FreeSurfer’s recon-all processing pipeline \cite{fischl2000measuring}, we obtained bilateral left and right segmentations of subcortical regions (13 total) and parcellations of cortical regions (33 total). We examined subcortical and cortical volumes, cortical surface area, and cortical thickness measures. 

\begin{figure*}
\centering
\includegraphics[width=\textwidth, height = 3in]{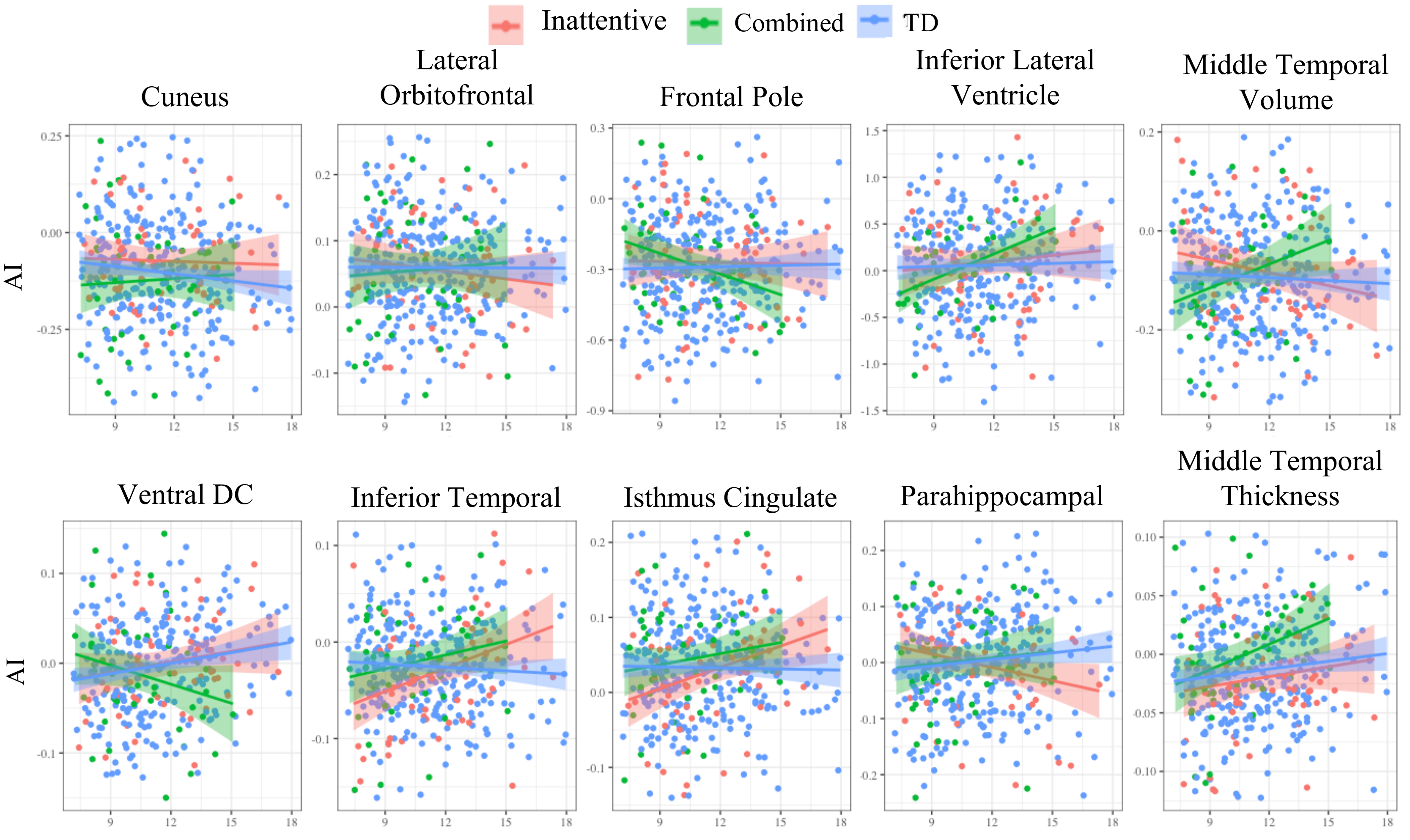}
%\end{center}
\caption{Plots with significant regions for the interaction of AI~(Y-axis) and age~(X-axis) across the presentations of ADHD Inattentive (red), ADHD Combined (green), and typically developing (TD; blue). An AI of zero indicates symmetry across both hemispheres, a negative AI indicates greater right laterality, and a positive AI indicates greater left laterality. Ventral DC = ventral diencephalon.}\label{fig:1}
\end{figure*}

\section{Statistical Analysis}
\label{sec:models}
\vspace{-5pt}

Asymmetry index (AI) is evaluated for each structure (i) and subject (j) across left (L) and right (R) hemispheres, where  $\mu$ is the average across the hemispheres \cite{guadalupe2017human,douglas2018hemispheric}:
\begin{equation}
\label{equation:AI}
AI\textsubscript{ij} = \frac{L\textsubscript{ij} - R\textsubscript{ij}}{\mu\textsubscript{ij}}
\end{equation}

A positive AI value indicates the left hemisphere is greater than the right; and conversely a negative AI value indicates the right hemisphere is greater than the left.

Using a linear mixed effect (LME) model, we define $Y$ as the dependent variable for a region-specific difference between the left hemisphere and right hemisphere (i.e., AI). We examined the anatomical metrics of volume, surface area, and thickness. The fixed-effect parameters in the model correspond to the covariates and factors which include: (1.) presentation group (ADHD-I, ADHD-C, and Typically-Developing or TD); (2.) age; (3.) interaction between age and presentation. To model variation across sites and account for correlation between observations within the same site, we used a site-specific random intercept in the model. The R-package {\tt lmertest} was used for fitting the models.

With presentation as a categorical variable with 3 levels, the binary indicator variables were the following: For the TD group, $D_{0i} = 1$ if the $i-$th participant is in the TD group and $D_{0i} = 0$ otherwise, ADHD-C as $D_{1i}$ and the ADHD-I as $D_{2i}$. A separate model examined only the binary variables of ADHD-I and ADHD-C for comparisons between them. Therefore, in this second model, for the ADHD-C group, $D_{0i} = 1$ if the $i-$th participant is in the ADHD-C group and $D_{0i} = 0$ otherwise, ADHD-I as $D_{1i}$.

We denote age of the subject $i$ to be $Age_i$. With $4$ sites available, we use $4$ binary indicator variables defined for Peking University ($S_{0i}$), Kennedy Krieger Institute ($S_{1i}$), New York University Child Center ($S_{2i}$) and the Oregon Health Science University ($S_{3i}$) and site-specific (deviation) zero-mean random intercepts, denoted $b_{0}, b_{1}, b_{2}$ and $b_{3}$. The following is the LME model: 
\[
Y_i = \sum_{s=0}^{3} b_s S_{si} + \sum_{d=0}^{2} ( \beta_{0}^{d} + 
\beta_1^{d} Age_i )*D_{di} + \epsilon_{i}  
\]
where $\epsilon_i$ is iid $N(0, \sigma_{\epsilon}^{2})$. There are separate regression curves for each presentation group (ADHD-I, ADHD-C and TD). Each presentation group's regression curve is also shifted vertically (upward or downward) depending on the site. Separate models for AI response variable $Y_i$ is defined for each of the anatomical metrics of volume, surface area, and thickness. The rationale for the interaction of age is due to its influence on brain asymmetry \cite{guadalupe2017human}. %We disregarded sex and handedness because of near ratios for sex in ADHD \cite{ramtekkar2010sex} and no significance in brain asymmetry for handedness in controls \cite{guadalupe2017human}.

\section{Results}
\label{sec:results}
\subsection{ADHD Presentations vs. Controls}
\label{ssec:adhdandcontrols}

\begin{figure*}[t!]
\centering
\includegraphics[width=\textwidth, height = 3in]{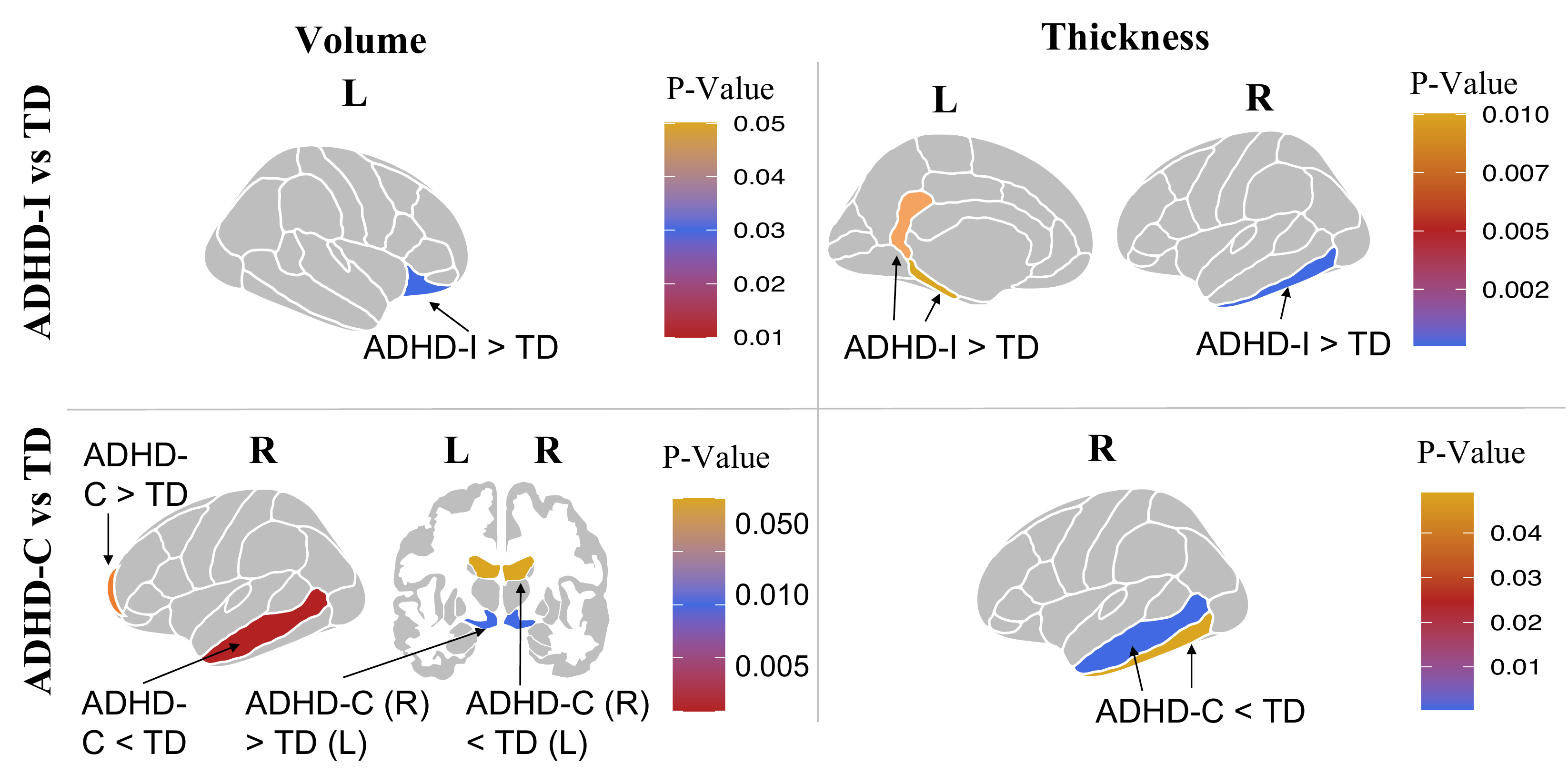}
%\end{center}
\caption{Significant regions ($p<0.05$) found between ADHD Presentations and Typically Developing (TD), with left (L) and right (R) laterality indicators, and greater than ($>$) or less than ($<$) AI. These included five volumetric regions (ADHD-I lateral orbitofrontal: $p = 0.034$; ADHD-C frontal pole: $p = 0.034$, inferior lateral ventricle: $p = 0.038$, middle temporal: $p = 0.008$, ventral diencephalon: $p = 0.010$) and five thickness regions (ADHD-I isthmus cingulate: $p = 0.008$, parahippocampal: $p = 0.009$, inferior temporal: $0.0002$; ADHD-C inferior temporal: $p = 0.049$, middle temporal: $p = 0.009$).}\label{fig:2}
\end{figure*}

\begin{figure*}
\centering
\includegraphics[width=\textwidth, height = 2.2in]{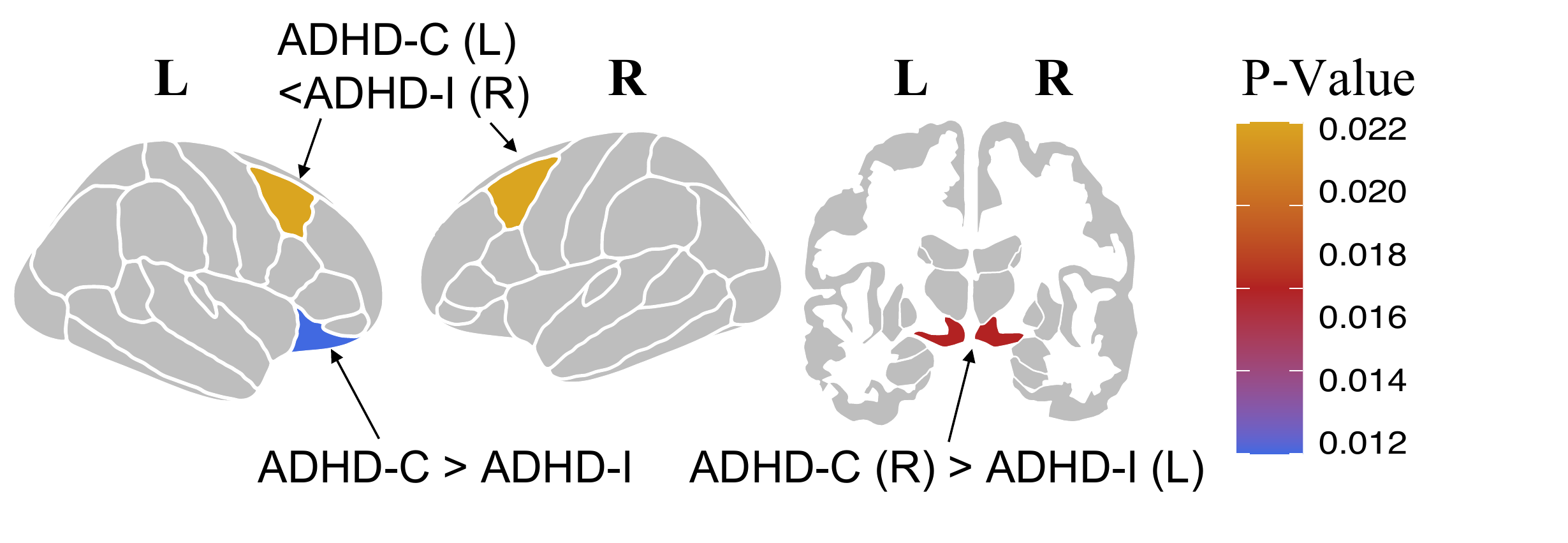}
\caption{ADHD-I vs ADHD-C
:~Three significant regions ($p<0.05$) found between ADHD-I and ADHD-C comparisons for volume (ventral diencephalon, $p = 0.018$), surface area (lateral orbitofrontal, $p = 0.012$), and thickness (caudal middle frontal, $p = 0.022$), with left (L) and right (R) laterality indicators, and greater than ($>$) or less than ($<$) AI.}\label{fig:3}
\end{figure*}

Analyzing the three categorical groups, we found five significant regions of AI in volume.~Volume AI was greater in ADHD-I youths compared to TD in the lateral orbitofrontal (${t[406] = 2.10, p = 0.036}$; left laterality) and its interaction with age ($t[406] = -1.97, p = 0.049$; ADHD-I less AI with increasing ages compared to TD), and interaction of age and cuneus (${t[402] = 2.30, p = 0.022}$; right laterality and ADHD-I less AI with increasing ages compared to TD). ADHD-C youths were significantly different in volume AI compared to TD in the following regions and interaction with age, respectively: frontal pole (${t[406] = 2.13, p = 0.034; t[406] = -2.08, p = 0.037}$; right laterality and ADHD-C increase AI with age compared to TD), inferior lateral ventricle ($t[341] = -2.08, p = 0.038; t[343] = 2.13, p = 0.034$; left laterality in TD, and increase AI and right to left in ADHD-C with increasing age), middle temporal ($t[406] = -2.67, p = 0.008; t[406] = 2.55, p = 0.011$; right laterality and ADHD-C decreasing AI with age compared to TD), and ventral diencephalon (${t[357] = 2.59, p = 0.010; t[357] = -2.77, p = 0.006}$;~left to right with increasing ages in ADHD-C, and right to left with increasing ages in TD). %Similar to lateral ventricle, disrupted water diffusion is found in ADHD-C right hemisphere \cite{hong2014connectomic}.

On the other hand, five significant thickness AIs were found comparing ADHD presentations with TD. Regions with greater AI in ADHD-I than the TD group along with the interaction of age, respectively, included the inferior temporal (${t[406] = -3.72, p = 0.0002; t[406] = 3.54}, p = 0.0005$; right to left with increasing ages in ADHD-I, right laterality in TD with increasing ages), isthmus cingulate (${t[407] = -2.68, p = 0.008; t[406] = 2.59, p = 0.010}$; left and increasing AI with age in ADHD-I), and parahippocampal (${t[406] = 2.62, p = 0.009}; t[406] = -2.69,p = 0.007$; left to right with increasing ages in ADHD-I, right to left with increasing ages in TD). Increased functional connections are found in posterior cingulate and parahippocampal \cite{anderson2014non}, supporting our results of increase AI from ADHD-I. ADHD-C differed from TD in the inferior temporal $(t[406] = -1.97, p = 0.049$; right laterality, increase AI with age in ADHD-C), middle temporal (${t[405] = -2.64, p = 0.009}$; overall decrease AI and right to left with increasing age in ADHD-C), and their interactions with age (respectively, (${t[406] = 2.02, p = 0.044; t[405] = 2.82}, p = 0.005$). Middle/inferior temporal BOLD fMRI increases in ADHD compared to TD \cite{spinelli2011variability}, whereas our results display increase (ADHD-I) and decrease (ADHD-C) AI compared to TD. We plot the interaction of age and AI significant regions across groups (Figure \ref{fig:1}), and display group significant regions (without interaction of age; Figure \ref{fig:2}).

\subsection{ADHD-I vs. ADHD-C}
\label{ssec:adhdiandadhdc}
When we considered a two-way comparison of our LME model to the ADHD presentations, we found few significant regions. The volume AI of the ventral diencephalon is less in ADHD-I (${t[101] = -2.41}, p = 0.018$) compared to ADHD-C. Similarly, greater structural connections in motor networks were found in ADHD-C \cite{saad2017regional}. When we examine interactions with age, there is a right to left shift from younger to older subjects from ADHD-I (${t[101] = 2.62, p = 0.009}$) whereas the reversal is shown in ADHD-C (i.e., left to right). The surface area of the lateral orbitofrontal appears to be left lateralized in both presentations (${t[108] = 2.56, p = 0.012}$), whereby older subjects have either a decrease in AI if they are ADHD-I (${t[108] = -2.40, p = 0.018}$) or an increase in AI if they are ADHD-C. Greater left orbitofrontal volume was found in ADHD-C previously \cite{plessen2006hippocampus}. Finally, thickness of the caudal middle frontal was significantly different between groups with increased AI in ADHD-I (${t[107] = 2.33, p = 0.022}$; left laterality) compared to ADHD-C (right laterality). Age and AI trends for significant regions are illustrated (Figure \ref{fig:1}). Figure \ref{fig:3} displays all significant AI regions between ADHD-I and ADHD-C.

\vspace{-5pt}
\section{Conclusion}
\label{sec:discussion}
The following are the novel findings: (1.) Our structural results support functional studies \cite{saad2020review} of disrupted motor networks in ADHD-C and cingulo-frontal networks in ADHD-I; (2.) Subnetwork regions of the ventral and dorsal default mode network (DMN) vary in structural asymmetry for all presentations of ADHD; (3.) Asymmetry differences in the temporal cortices vary for all subtypes and should be examined further; (4.) Neurodevelopment trajectories suggest certain regions normalize in hemispheric asymmetry. This normalization could be causal or compensatory in nature.

\section{Acknowledgments}
\label{sec:acknowledgments}
We appreciate the efforts of The Neuro Bureau and Virginia Tech’s ARC team for providing the ADHD-200 preprocessed data repository. We give profound thanks to the King Abdullah University of Science and Technology for providing our team with the baseline funds to conduct this work. Last, we thank the KAUST Biostatistics Group and Dr. Aritra Dutta for their feedback and suggestions. The authors acknowledge the support provided in part by the NARSAD Young Investigator Grant and the K01 NIH (1MH110645).

\vspace{-5pt}

\bibliographystyle{IEEEbib}
{\small
\bibliography{strings,refs}
}

\end{document}